\documentstyle[multicol,aps,harvard,floats]{revtex}

\newcommand{\beq}{\begin{equation}}
\newcommand{\eeq}{\end{equation}}
\citationstyle{dcu}

\begin{document}

%\special{papersize=8.5in,11in}
\title{Waves in Open Systems: Eigenfunction Expansions}

\author{E. S. C. Ching${}^{(1)}$, P. T. Leung${}^{(1)}$, 
W. M. Suen${}^{(1,2)}$, S. S. Tong${}^{(1)}$ and K. Young${}^{(1)}$}

\address{${}^{(1)}$Department of Physics, 
The Chinese University of Hong Kong, Hong Kong, China\\
${}^{(2)}$McDonnell Center for the Space Sciences,
Department of Physics, Washington University, St. Louis, Missouri 63130, USA}

\date{\today}

\maketitle

\begin{abstract}
An open system is not conservative because energy can escape to
the outside. An open system by itself is thus not conservative.  
As a result, the time-evolution operator is not hermitian in the 
usual sense and the eigenfunctions
(factorized solutions in space and time) are no longer normal modes
but quasinormal modes (QNMs) whose frequencies $\omega$ are complex. 
QNM analysis has been a powerful tool for investigating
open systems. Previous studies have been mostly system specific,
and use a few QNMs to provide approximate descriptions. 
Here we review recent developments which aim at
a unifying treatment.
The formulation leads to a mathematical structure in close
analogy to that in conservative, hermitian systems. Many of
the mathematical tools for the latter can hence be transcribed.
Emphasis is placed on those cases in which the QNMs form a complete
set for outgoing wavefunctions, so that in principle all the QNMs
together give an exact description of the dynamics. 
Applications to optics in microspheres
and to gravitational waves from black holes are reviewed, 
and directions for further development are outlined.

\end{abstract}

%==========================================================
%\newpage
\begin{multicols}{2}

%\tableofcontents

\pagestyle{myheadings}
\markright{\today / Waves in open systems: Eigenfunction expansion
/ Draft}

%==========================================================

%==========================================================
%file sect1

\section{Introduction}
\label{sect:intro}

%==========================================================
\subsection{Physical motivation}
\label{subsect:phys}

Many concepts in physics rely on the idea of the normal modes (NMs) of
a conservative system.  Thus one speaks of energy eigenstates,
molecular orbitals, energy bands, transitions between states and
excitation energies (at the first quantized level), or of propagators
in a Feynman diagram (at the second quantized level).  These concepts
for {\it quantum} and {\it interacting} (i.e., nonlinear) systems are
rooted in eigenfunction expansions in terms of the NMs of {\it
classical} and {\it linear} systems; e.g., the photon propagator in
QED depends on the free classical electromagnetic (EM) field being
expandable in plane waves.  Such expansions are possible because
conservative systems are associated with hermitian operators.

On the other hand, if energy can escape to the outside,
the system would be open and nonconservative,
the associated mathematical operators are not 
hermitian in the usual sense. The eigenfunctions
are then quasinormal modes (QNMs) with complex frequencies $\omega$.
Figure 1a shows schematically the EM spectrum observed outside a linear
optical cavity of length $a$, due to emission by a broad-band 
source, e.g., fluorescent dye molecules.  
The resonances, spaced by $\Delta \omega \approx \pi c/a$
(where $c$ is the speed of light), are QNMs, which are
characteristic of the cavity rather than of the source.
The resonance width $\gamma$ is determined
(in the absence of absorption) by the amount of leakage.
In the limit of zero leakage, these 
QNMs reduce to the NMs of a conservative cavity.
It would be both interesting and useful if the physics of open systems
could be discussed in terms of the discrete QNMs, in other words,
an eigenfunction expansion for the dynamcis.

This task is nontrivial because there are very few known results
for eigenfunction expansion in
nonconservative, non-hermitian systems.  Yet the formulas we show
in this article will look extremely familiar --- and that is precisely the point
of this paper: the familiar formalism, with very little alteration, apply to 
a wide class of these systems.

The problem of EM waves escaping from an open cavity is brought into 
sharp focus by advances in cavity QED:
in Fabry-Perot cavities \cite{fab1,fab2}, in superconducting
microwave cavities \cite{mic1,mic3}, in semiconductor
heterostructures \cite{semi1}   
and in microspheres which confine
glancing rays by total internal reflection \citeaffixed{mie2,sando}{see,
e.g.,}.
First, the spectrum is dominated by the resonances.
Figure 1b shows an experimental spectrum observed from a
microsphere, similar in essence to Figure 1a.
Second, the decay rate of an excited atom or molecule
is enhanced (suppressed) if the emitted
radiation falls on (away from) the resonances
\cite{mic1,fab1,fab2,barnes} --- 
the atom or the molecule of dimension $\mbox{$\sim 0.1 $ nm}$ ``knows"
about its environment, on a
scale of $\mbox{$\sim 1 \, \mu$m}$, even before any photon is emitted.

Gravitational waves can be produced by matter falling into a
black hole, and may soon
be observed \citeaffixed{abram}{see, e.g.,}.
The curvature of the intervening space 
forms an effective potential
akin to a cavity, from which the gravitational
waves eventually escape.
Numerical simulations
\cite{vish,det,smarr,stark,anninos} show that
the amplitude is dominated, at intermediate times, by a ringing
signal (Figure 2) $\sum a_j e^{-i\omega_j t}$, where $j$ labels the QNMs,
and each complex frequency $\omega_j$ is 
characteristic of the background geometry 
rather than of the emitting source.
 
In these and many other examples,
the resonances or QNMs often dominate.
But how complete are the QNMs? 
In mathematical terms,
this means, first of all, whether any function $\phi(x)$
can be expanded in terms of the QNMs $f_j(x)$:

%--------------------------------------------------------------------------------------------------expand1
\beq
\phi(x) = \sum_j a_j f_j(x)
\label{eq:expand1}
\eeq

\noindent
More importantly, do the resonances represent
the dynamics exactly:

%--------------------------------------------------------------------------------------------------expandt
\beq
\Phi(x,t) = \sum_j a_j f_j(x) e^{-i\omega_j t}
\label{eq:expandt}
\eeq

\noindent
for all $t \ge 0$? These expansion should make no reference
to the environment or external bath.
One would wish to establish conditions under which these expansions are valid,
and in circumstances where they are not, to 
characterize the ``remainder".

A projection formula, i.e., some sort of inner product, is needed to obtain the
coefficients $a_j$ from $\phi(x)$.
Then, initial value problems become formally
trivial: take the initial data, project out $a_j$, and
evolve by $a_j \mapsto a_j e^{-i\omega_j t}$.
Moreover, to the extent that similarities can be established with the conservative case,
one should be able to transcribe
the tools of mathematical physics,
and to seek a parallel formalism ---
including, e.g., second quantization.
These developments should lead to a unifying view of QNMs in different systems.

This article discusses recent progress towards these goals.

Open systems are a special class of dissipative systems, 
which have been studied by many authors, including
\citeasnoun{feyn},
\citeasnoun{uller},
\citeasnoun{cald}.
In all cases, the bath degrees of freedom are first included, and then
eliminated from the path integral or equations of motion.

Similarly, an open cavity
(linear dimension $a$) can be embedded in a ``universe" (linear dimension
$\Lambda \rightarrow \infty$), and the totality is 
a conservative system with the modes of the universe (MOU) forming
a continuum ($\Delta \omega \propto \Lambda^{-1} \rightarrow 0$).
A rigorous theory of
lasing in 1-d cavities has been developed using the MOU 
\cite{lang1,lang2} and
the ideas can be generalized to
higher dimensions, e.g., optics in microspheres
\citeaffixed{ching1,ching2}{see, e.g.,},
including nonlinear optical processes
such as spontaneous and stimulated Brillouin scattering
\cite{bs,sbs}.

Since each eigenfunction or QNM is a resonance, the
expansions (\ref{eq:expand1}) and (\ref{eq:expandt})
in effect expresses the idea of resonance domination.
In cavity QED, 
resonance domination was first discussed by
\citeasnoun{purcell}.
He proposed that in the Fermi golden rule, the
the density of states per unit volume
$\rho_0(\omega) = \omega^2/(\pi^2c^3)$ should be replaced by
$\rho(\omega) \sim D/(2\gamma V)$
for a $D$-fold degenerate QNM of width $\gamma$ in a cavity of
volume $V$. This leads to an 
enhancement factor of $K = \rho/\rho_0 \sim (1/8\pi)DQ (\lambda^3/V)$ 
for spontaneous emission, where $\lambda$ 
is the wavelength of light emitted
and $Q$ is the quality factor of the cavity.  Purcell's argument
can be made more precise using
the MOU \cite{ching1,ching2}:
the photon modes are re-distributed
--- enhanced at the resonance and 
depleted away from the resonances.

These works, in company with much earlier works in
metastable states in nuclear physics \cite{gamow,siegert} typically
use {\em some} resonances for an {\em approximate} description.
Our focus will be on the use of {\em all} the discrete resonances to obtain, if possible,
an {\em exact} description --- and where this fails, to characterize the
remainder.

%==========================================================

\subsection{The mathematical problem}
\label{subsect:math}

Waves in open systems may be governed by the Schr\"odinger equation
(e.g., nuclear physics), the wave equation (e.g., optics) or the Klein-Gordon (KG)
equation with a potential $V$ (e.g., each angular momentum sector of linearized
gravitational waves \cite{chand}).  The wave equation can be mapped exactly
into the KG equation, and as far as time-independent poblems are concerned,
the KG equation can be mapped into the Schr\"odinger equation by relabeling
$\omega^2 \rightarrow \omega$.  Thus, many properties are similar (although
we shall also highlight the key differences), and we shall
focus on the scalar wave equation, on 
a certain domain ${\cal R}$,

%--------------------------------------------------------------------------------------------------wave3d
\beq
D \, \Phi( {\bf r} , t ) \equiv \left[ \rho( {\bf r} ) \partial_t^2
- \nabla^2 \right] \Phi ( {\bf r} , t) = 0
\label{eq:wave3d}
\eeq

\noindent
This describes the scalar model
of EM ($\rho =$ dielectric constant, and henceforth
$c=1$)
or elastic vibrations ($\rho =$ density).  
In this article, we shall for the most part consider the 1-d 
version ($\nabla^2 \mapsto \partial_x^2$),
restricted to a half line $0 \le x < \infty$;
the ``cavity" is the interval ${\cal R} = [0,a]$.
A node is imposed at the origin (a totally reflecting mirror),
and waves escape through the point $x=a$ (a partially transmitting mirror)
to the rest of the ``universe" in $(a, \infty)$.  
This model describes a laser cavity \cite{lang1},
as well as gravitational radiation from a stellar object \cite{price},
or the transverse vibrations of a string \cite{dekker4,laistring}.

For conservative systems, one usually
imposes $\Phi =0$ on the boundary
$\partial {\cal R}$ ($x=a$ in the 1-d case)
\footnote{The Dirichlet condition can be replaced
by the Neumann condition.}.
The eigenfunctions or NMs are factorized solutions
$\Phi( {\bf r}, t) = f( {\bf r} ) e^{-i\omega t}$.
The nodal condition on $\partial {\cal R}$ implies that ${\cal R}$ 
is {\it closed},
so that energy cannot escape.  Mathematically, the operator $-\nabla^2$ is
hermitian; thus $\omega$
is real and the eigenfunctions $\{ f \}$
form a complete orthogonal set. 

In contrast, if $\Phi$ satisfies the outgoing wave condition on $\partial {\cal R}$;
this defines an {\em open} system. The eigenfunctions, labeled by an index $j$, satisfy
(in 1-d)
$\partial_x^2 f_j (x) = -\omega_j^2 \rho(x) f_j(x)$,
with $\mbox{Im } \omega_j < 0$.
Note a doubling of modes compared with the conservative
case:
If $\omega_j$ is an eigenfrequency, then so is
$\omega_{-j} \equiv -\omega_j^*$, but
since $\omega_{-j}^2 \neq \omega_j^2$, 
$f_{-j}(x)$ and $f_j(x)$ are linearly independent.
The question then is whether these $\{ f_j \}$ are complete
in the sense of (\ref{eq:expand1}) and (\ref{eq:expandt}).
 % Introduction --- \label{sect:intro}
%==========================================================
%file sect2

\section{Completeness}
\label{sect:comp}

Even though QNMs do not always capture all the physics,
it is still useful to first establish the baseline case
in which the QNMs are complete 
on ${\cal R} = [0,a]$.
This holds provided two
conditions are satisfied:
(a)  $\rho (x)$ has a step or stronger discontinuity
at $x=a$ (discontinuity condition);
(b)  $\rho(x)=1$ (or other constant value)
for $x>a$ (``no-tail" condition).
These two conditions can be understood heuristically.
A discrete representation of $\phi(x)$ such as (\ref{eq:expand1})
could at best be valid over a finite interval, for which a 
discontinuity provides
a natural demarcation.  Moreover, a dynamical expansion
such as (\ref{eq:expandt}) would require that waves are not
affected by the outside $(a,\infty)$; this can only be the case if $\rho$
is trivial outside, and thus outgoing waves are not scattered back.

We now sketch the main elements of the proof
of completeness, which will also provide the framework for
understanding other contributions when the QNMs are not complete.
The Green's function $G(x,y;t)$, defined by 
$D G(x,y;t) = \delta(x-y) \delta(t)$, with
$G(x,y;t)=0$ for $t \leq 0$. 
Its fourier transform 
${\tilde G}(x,y;\omega)$
is given explicitly as
${\tilde G}(x,y,\omega) = f(\omega,x) g(\omega,y) / W(\omega)$
for $0 \leq x \leq y$.  Here $f$ and $g$ are homogeneous solutions
to the time-independent wave equation at frequency $\omega$, 
with $f$ satisfying the left nodal boundary condition ($f(\omega,0) = 0$)
and $g$ satisfying the right outgoing wave boundary condition
($g(\omega,x) \propto e^{i\omega x}$ for $x \rightarrow \infty$).
Their Wronskian is $W$, and the combination $fg/W$
is independent of normalization convention.

For $t \ge 0$, $G(x,y;t)$ is evaluated by the inverse Fourier transform
with the contour of integration
closed by a large semicircle in
the lower half $\omega$-plane. In general, there are three different
contributions.

\noindent
{\em Large semicircle --- prompt response}

The contribution along the
large semicircle ($|\omega| \rightarrow \infty$) gives
the short time or prompt response.  For large $t$, the factor
$e^{-i\omega t}$ provides sufficient damping in the lower half
$\omega$-plane to control the asymptotic behavior and
it can be shown 
\cite{bachelot,lly1} that
the prompt response always vanishes for $t > t_p(x,y)$,
where $t_p(x,y)$ is the geometric optics transit time   
between $x$ and $y$.

If the discontinuity condition is satisfied,
one has a stronger result:
by a WKB estimate,
this contribution vanishes for all $ t \ge 0$.  It is
natural that the $|\omega| \rightarrow \infty$ behavior
should be sensitive to short spatial length scales,
in particular a discontinuity.

\noindent
{\em Singularities in $g$ --- late-time tail}

The function $f(\omega,x)$ is obtained by integrating 
the time-independent wave equation,
in which $\omega$ appears analytically, through a {\em finite}
distance from $0$ to $x$. Crudely speaking, this is a finite combination
of analytic functions of $\omega$, so $f(\omega,x)$ is guaranteed to be
analytic in $\omega$ \cite{newton}. On the other hand, 
$g(\omega,x)$ could have discontinuities in $\omega$
(typically a cut on the negative $\mbox{Im } \omega$ axis),
because it is obtained by integrating through an infinite
interval starting from $x = \infty$.

However, if the ``no tail" condition is satisfied, then one can impose
the right boundary condition for $g(\omega,x)$ at $x=a^+$;
one then again integrates only through a finite distance, thus
guaranteeing the analyticity of $g$ as well, and removing this contribution.

\noindent
{\em Zeros of W --- QNM contributions}

Finally, there are contributions from the zeros of $W(\omega)$.
At a zero $\omega_j$ of $W$, the functions $f$ and $g$ are linearly
dependent: $f_j(x) \equiv f(\omega_j , x) = C_j g(\omega_j , x)$; 
thus $f_j$ satisfies
{\em both} the left nodal and the right outgoing-wave boundary conditions, 
and is an eigenfunction or QNM with eigenfrequency $\omega_j$.

In general, all three contributions are present, and correspond
nicely with the main features of the time domain signal
\cite{taillett,tail}, as shown in
the spacetime diagram in Figure 3.
The contribution from the semicircle in the $\omega$-plane leads to the
initial transients --- a wave propagating directly from the source point $y$ to
the observation point $x$ (rays $1a, 1b$).
The zeros of $W$ give rise to the QNM ringing at intermediate times,
which are caused by repeated scattering from the potential (ray $2$).
The cut in the $\omega$-plane, especially its tip near $\omega=0$, then
gives the late-time tail: 
waves propagate from the
source point $y$ to a distant point $x'$, are scattered by
$\rho(x')$, and return to the observation
point $x$ (ray $3$).

We now focus on the case where the discontinuity and ``no-tail"
conditions are satisfied.
At each zero $\omega_j$, the residue is related to $dW(\omega_j)/d\omega$, which can
be evaluated using the defining equations for $f$ and $g$ to be

%--------------------------------------------------------------------------------------------------norm
\begin{eqnarray}
-\frac{1}{C_j} \frac{dW(\omega_j)}{d\omega}
&=&  2 \omega_j \int_0^{a^+}\mbox{\Large{}} \rho(x) f_j(x)^2 dx
+ i f_j(a^+)^2 \nonumber \\
& \equiv & \langle f_j | f_j \rangle
\label{eq:norm}
\end{eqnarray}

\noindent 
Using the normalization (and phase) convention 
$ \langle f_j | f_j \rangle = 2\omega_j$, one then obtains

%-------------------------------------------------------------------------------------------------grep
\beq
G(x,y;t) = \frac{i}{2} \sum_j \frac{1}{\omega_j} f_j(x) f_j(y)
e^{-i\omega_j t}
\label{eq:grep}
\eeq
 
\noindent
The initial conditions $G(x,y;t=0)=0$,
$\rho(x) \partial_t G(x,y;t=0) = \delta(x-y)$
lead to\footnote{Subject to the validity of
term-by-term differentiation and the limit $t \rightarrow 0$.}
 
%--------------------------------------------------------------------------------------------------id1
\begin{eqnarray}
\frac{i}{2} \sum_j \frac{1}{\omega_j} f_j(x) f_j(y) &=& 0 
\label{eq:id1} \\
\frac{ \rho(x) }{2}\sum_j  f_j(x) f_j(y) &=& \delta(x-y)
\label{eq:id2}
\end{eqnarray}

\noindent
for $x,y \in {\cal R}$.
The identity (\ref{eq:id2}) has an obvious analog in conservative
systems; the factor $1/2$ accounts for the doubling of modes.
On the other hand (\ref{eq:id1}) 
becomes vacuous in the limit $\epsilon \rightarrow 0$.

The heart of the eigenfunction expansion is contained in
(\ref{eq:grep}), (\ref{eq:id1}) and (\ref{eq:id2}).
The identity (\ref{eq:id1}) obviously leads to (\ref{eq:expand1}),
while (\ref{eq:grep}) will be seen to lead to (\ref{eq:expandt}). % Completeness --- \label{sect:comp}
%========================================================== 
%file sect3 

\section{Time-independent problems} 
\label{sect:timeindep} 

This simplest and most useful applications concern
time-independent problems, especially time-independent perturbation theory.

%========================================================== 
\subsection{Perturbation} 
\label{subsect:pert1}

Let 
$\rho (x) =  \rho_0 (x) + \mu V(x)$, where $|\mu |\: \ll 1$,
and $V(x)$ is nonzero only in the cavity.  The system with $\rho_0 (x)$ is
exactly solvable and its QNMs form a complete set for $x \in [0,a]$
in the sense of (\ref{eq:grep}), (\ref{eq:id1}) and (\ref{eq:id2}).
Thus, it is straightforward to develop a {\em discrete}
representation for the frequency shifts
$\omega _j = \omega ^{(0)}_j + \mu  \omega ^{(1)}_j
+ \mu ^{2} \omega ^{(2)}_j + \cdots $ :

%--------------------------------------------------------------------------------------------------shift1a
\beq
\omega ^{(1)}_j = - {\omega ^{(0)}_j \over 2} V_{jj}
\label{eq:shift1a}
\eeq

%--------------------------------------------------------------------------------------------------shift2a
\beq
\omega ^{(2)}_j = {\omega ^{(0)}_j \over 4}\Bigl\{2(V_{jj})^2 + 
\sum^{}_{i\neq j}\frac{ V_{ji} V_{ij} \omega ^{(0)^2}_j} 
{ \omega ^{(0)}_i(\omega ^{(0)}_j - \omega ^{(0)}_i)}  \Bigr\} 
\label{eq:shift2a}
\eeq

\noindent
etc., where the matrix elements are
$V_{ji} = \int_0^a f_j^{(0)}V f_i^{(0)} dx$
These complex 
formulas give the shifts in the resonance positions and the
widths --- thus they contain twice the information compared
with their superficially similar conservative counterparts.  The wavefunctions can be calculated
in a parallel way \cite{lly2}.

The first-order correction has been known for a long time for 
the Schr\"{o}dinger equation, but here the normalization 
condition (see (\ref{eq:norm})) is stated through
a surface term rather than via regularization \cite{zel}, which is
less convenient computationally.  The first-order result has been generalized to the EM 
case \cite{per1}, and in fact
applies also to systems without discontinuities.

To illustrate,
let $\rho(x) = 1 + n^2 + \mu \Theta(x-b)$ for $x<a$ 
and $\rho(x) = 1$ for $x>a$ (i.e. a ``dielectric rod"
with a ``bump").
Figure 4 shows the QNM positions in the $\omega$-plane
for $n^2 = 4.0$, $\mu=0.4$, compared with $\mu=0$. 
The two are quite different, in that for the 
former case $\mbox{Im $\omega_j $}$ are oscillatory in the mode number $j$. 
The perturbative results based on (\ref{eq:shift1a}) and (\ref{eq:shift2a})
are also shown in Figure 4; although these formulas
mimic the case of {\em real \/} frequencies for conservative systems,
they nevertheless give the correct shifts, even for the imaginary parts.
The sum over intermediate states is discrete, at a spacing $\Delta \omega 
\sim \pi / a$; thus the effective expansion parameter is 
$\mu / |\Delta \omega| \sim \mu a / \pi$, a perspective
not apparent in the MOU approach. 

%==========================================================
\subsection{Physical examples}
\label{subsect:microdroplet}

Some interesting applications relate to optics;
the interface between two media
naturally gives a discontinuity, and vacuum outside
the system naturally satisfies the ``no tail" condition.
Consider a microsphere of radius $a$ and  
refractive index $n$.  Glancing rays suffer total internal 
reflection and are confined, but
evanescent waves cause some leakage, and this makes the system an open one.
In practice, experiments are often done on droplets falling in air, 
slightly oblate (ellipticity $e \sim 10^{-3} - 10^{-2}$) due to viscous drag. 
The breaking of spherical symmetry lifts the degeneracy 
among the $2l+1$ members of a multiplet.  This problem 
has been treated by brute force numerical calculation \cite{hill}, 
but the perturbative formalism allows an analytic expression \cite{per1} 
 
%--------------------------------------------------------------------------------------------------split 
\beq 
\frac{\omega_m^{(1)}}{\omega^{(0)}} 
= -\frac{e}{6} \left[ 1 - \frac{3m^2}{l(l+1)} \right] 
\label{eq:split}  
\eeq 
 
\noindent 
in which $m$ is the azimuthal quantum number, and
$e = (r_p - r_e)/a$, where $r_p$ and $r_e$ are respectively the polar 
and equatorial radii, and $a = (r_p r_e^2)^{1/3}$.
This perturbative formula has been used to interpret 
spectroscopic measurements and to determine the ellipticity $e$,
either by measuring the splitting between two neigboring
lines  ($m$ and $m+1$) \cite{split1}, or by determining the
total spread of the multiplet (from $m=0$ to $m= \pm l$) 
\cite{split2}.
Two other developments along this line should be mentioned. 
 
In one interesting experiment \cite{arnold}, 
a quadrupole field $E_Q$ imposed on a levitated droplet caused 
a spheroidal distortion with an 
ellipticity $e \propto E_Q/s$, where $s$ is the surface tension. 
The splitting of the multiplet was determined spectroscopically 
and the ellipticity $e$ was then found from (\ref{eq:split}),
allowing $s$ to be determined,
down to length scales of tens of $\mu$m.  (Hitherto, surface tension 
is known only macroscopically, at length scales of say 1 mm or more.)
 
In another experiment \cite{prec}, light is launched into  
a slightly oblate microdroplet,
propagating along great circles inclined 
at an angle $\theta$ to the equatorial plane.  In photon 
language, the angular momentum ${\vec L}$ satisfies 
$m/l = L_z/L = \cos \theta$, and
precesses at the group angular velocity 
$\Omega = d \omega_m/ dm  
= \omega |e| m / [l(l+1)] $.  The precession is observed as 
variations in intensity when a 
particular spot at an angle $\theta$ is observed.  The observed 
period of typically 50 ps then allows the ellipticity to be determined. 

On a very different length scale, the perturbation results
have also been applied to gravitational waves.
The QNM spectrum of linearized gravitational waves propagating 
on a Kerr or Schwarzschild background
has been extensively studied analytically \cite{chand} 
and numerically \cite{vish,det,smarr,stark}, and their complex frequencies 
are known in detail \cite{leaver1,guinn,leaver2,andersson,nollert}.
However, black holes are perturbed by interactions with its 
surroundings (e.g., a massive accretion disk).  The shifts may provide
a probe of the intervening spacetime curvature.
This situation is obviously one to which perturbation theory can be applied.

The first-order perturbation theory is expressed 
by the KG analog to (\ref{eq:shift1a}), 
further generalized to allow for a ``tail"  
\cite{dirtybh}.\footnote{With a ``tail", the QNMs are 
not complete, but it is not surprising that this 
does not matter for the first-order shift.} 
We here present some results for a Schwarzschild black hole perturbed
by a thin static shell of matter \cite{dirtybh}.
Figure 5 shows, in the complex $\omega$-plane,
the locus of the lowest QNM for $l=1$ scalar waves
propagating on a Schwarzschild background, as a light shell
($0.01$ times the mass of the hole) is placed at different positions $r_s$.
The first-order perturbative
results (dashed lines) capture the qualitative features of the
exact results (solid lines).

 % Time-independent problems --- label{sect:timeindep}
%==========================================================
%file sect4

\section{Dynamics}
\label{sect:timedep}

Time-dependent problems require
two independent sets of initial data:
$\phi \equiv \Phi(x,t=0)$ 
and ${\hat \phi} \equiv \rho(x) \partial_t \Phi (x,t=0)$.
\footnote{These are assumed to vanish at infinity,
representing waves emitted at a finite time in the past.}
One is thus led to consider
the {\em simultaneous} expansion of a {\em pair} of functions 
$(\phi, {\hat \phi})$ 

%--------------------------------------------------------------------------------------------------expand2 
\beq 
\pmatrix{ \phi(x) \cr {\hat \phi}(x) \cr } =  
\sum_j a_j 
\pmatrix{ 1 \cr -i \omega_j \rho(x) \cr } f_j(x) 
\label{eq:expand2} 
\eeq 
 
\noindent 
using the {\it same} coefficients $a_j$'s 
for both components.   
The factor $\rho(x)$ 
turns the second component into the conjugate momentum.
The use of two
components is natural \cite{fesh,unruh,ford}, but here the outgoing wave condition
\footnote{The last condition is specified  
at $a^+$ because the system may have discontinuities at $x=a$.}
${\hat \phi} (x=a^+) = - \phi ' (x=a^+)$ 
links them in a novel way.

Compared with the case of NMs, twice the degrees of freedom ($a_j$ and $a_{-j}$) are used to  
satisfy twice the conditions ($\phi$ and ${\hat \phi}$).
In the conservative limit,
$\omega_{-j}^2 =  \omega_j^2$ and $f_{-j}(x) = f_j(x)$.   
Thus, upon grouping pairs of terms,
$\phi(x) = \sum_{j>0} b_j f_j(x)$, 
${\hat \phi}(x) = \rho(x) \sum_{j>0} c_j f_j(x)$, where 
$b_j = a_j + a_{-j}$, and 
$c_j = -i\omega_j (a_j - a_{-j})$. 
So in this limit, the simultaneous expansion of two functions 
$(\phi, {\hat \phi})$ using the full set of NMs $j= \pm1, \pm2, \cdots$ is 
equivalent to the separate and independent expansion of $\phi$  
and ${\hat \phi} / \rho$
using only the NMs $j=1,2, \cdots$. 

The solution to the initial value problem is then

%--------------------------------------------------------------------------------------------------dyn 
\beq 
\Phi (x,t)  
= \int_0^{\infty} \left[ G(x,y;t) {\hat \phi} (y) + \partial _t G(x,y;t)  
\rho(y) \phi (y) \right] dy 
\label{eq:dyn} 
\eeq 
 
\noindent 
From (\ref{eq:dyn}), one can then obtain (\ref{eq:expand2}) and
also (\ref{eq:expandt}) with 
$a_j$ given by 
 
%--------------------------------------------------------------------------------------------------aj 
\begin{eqnarray}
a_j = \frac{i}{2 \omega_j}  \bigg\{ \int_0^{a^+} 
&&\left[ f_j(y) {\hat \phi}(y) + {\hat f}_j(y) \phi(y) \right] dy  
\nonumber \\
&& + \, f_j(a) \phi(a)  \bigg\} 
\label{eq:aj} 
\end{eqnarray}
 
\noindent
All reference to the outside of the cavity
is removed, because the integral on $(a,\infty)$ can be collapsed to
the surface term, by making use of the outgoing condition on the
initial data and the retarded condition on $G$.

The elimination of the outside is the crucial step in
obtaining a self-contained description of the cavity.

The projection formula (\ref{eq:aj}) suggests the definition of a generalized inner 
product, and with it a linear space structure 
analogous to that for conservative systems \cite{twocomp1,twocomp2}.
Consider the linear space of outgoing function pairs $ (\phi, {\hat \phi})$,
denoted as ket vectors
$| \mbox{\boldmath $\phi$} \rangle = (\phi(x) , {\hat \phi}(x) )^{\rm T}$.
For a QNM,
${\hat f}_j(x) = -i\omega_j \rho(x) f_j(x)$,
where $\omega_j$ is the eigenvalue.
Time-evolution can be written in Schr\"odinger form
$\partial_t | \mbox{\boldmath $\phi$} \rangle \, = \,  
-i \cal{H} | \mbox{\boldmath $\phi$} \rangle$,
where  
 
%--------------------------------------------------------------------------------------------------htwocomp 
\beq 
\cal{H} = \mbox{$i$} \pmatrix{ 0 & \rho(x)^{-1} \cr \partial_x^2 & 0 } 
\label{eq:htwocomp} 
\eeq 
 
\noindent 
The first component of the Schr\"odinger equation reproduces the identification 
of ${\hat \phi}$ as $\rho(x) \partial_t \Phi$. 
The eigenfunctions, or QNMs, can now be defined simply by 
$\cal{H} \, | \mbox{\boldmath $f_j$} \rangle =  
\mbox{$\omega_j$} \, | \mbox{\boldmath $f_j$} \rangle$.

The projection (\ref{eq:aj}) suggests a generalized 
inner product between two vectors 
$| \mbox{\boldmath $\phi$} \rangle$ and 
$| \mbox{\boldmath $\psi$} \rangle$:

%--------------------------------------------------------------------------------------------------inner 
\beq 
\langle \mbox{\boldmath $\psi$}| \mbox{\boldmath $\phi$} \rangle  \,   =   \, 
i \left[ \int_0^{a^+ } ( \psi {\hat \phi} + {\hat \psi} \phi ) dx \, + \, \psi(a) \phi(a) \right] 
\label{eq:inner} 
\eeq 
 
\noindent 
which is symmetric and linear in both the bra and ket vectors (rather than conjugate 
linear in the bra vector).
The inner product of a QNM  
with itself agrees exactly 
with the generalized norm (\ref{eq:norm}).
Moreover, the coefficients (\ref{eq:aj}) for the eigenfunction expansion 
can now be written compactly as 
 
%--------------------------------------------------------------------------------------------------ajcomp 
\beq 
a_j = \frac{\langle \mbox{\boldmath $f$}_j  
| \mbox{\boldmath $\phi$} \rangle} 
{\langle \mbox{\boldmath $f$}_j  
| \mbox{\boldmath $f$}_j \rangle}  
= \frac{1}{2 \omega_j} \langle \mbox{\boldmath $f$}_j  
| \mbox{\boldmath $\phi$} \rangle 
\label{eq:ajcomp} 
\eeq 
 
Under the generalized inner product, the Hamiltonian is
symmetric: 
$\langle \mbox{\boldmath $\psi$} | \mbox{ {\large \{}} \cal{H} | \mbox{\boldmath $\phi$}
\rangle \mbox{ {\large \}}} = 
\langle \mbox{\boldmath $\phi$} | \mbox{ {\large \{}} \cal{H} | \mbox{\boldmath $\psi$}
\rangle \mbox{ {\large \}}} \equiv 
\langle \mbox{\boldmath $\psi$} | \cal{H} | \mbox{\boldmath $\phi$} \rangle$.
In the proof of this statement, there is an integration by parts;
surface terms are incurred because the functions do not vanish
at the end-points, but these are compensated exactly by the
surface terms in the definition of the inner product.

There is thus an almost complete parallel with conservative systems, and
the mathematical structure is in place to carry over  
essentially all the familiar tools based on eigenfunction expansions.   
For example, the symmetry of ${\cal H}$ (similar to the hermitian
property in the conservative case) immediately leads to the
orthogonality of the QNMs and hence the uniqueness of the expansion
(\ref{eq:expand2}).
The only exception is the lack of a positive-definite norm, and with it  
a simple probability interpretation --- hardly surprising since  
probability (or energy) is not conserved in ${\cal R}$. 
For example, the perturbation theory can now be obtained by an almost
trivial transcription of textbook derivations.

The formalism generalizes to a full line
($-\infty < x < \infty$); the expansion is valid in an interval 
$[a_1,a_2]$, provided  
(a) there are at least two 
discontinuities, with the extreme ones at $a_1$ and $a_2$;  
(b) there is  
``no tail" on either side, and (c) the outgoing condition is imposed at 
the two ends $a_1^-$ and $a_2^+$.  
Much the same formalism applies to the KG equation.
Discontinuities must be present in the potential $V(x)$,
and the ``no tail" condition is that $V(x)=0$ on the ``outside"
\cite{kgcomplett,kgcomp}.
Absorption can also be included;
e.g., in the scalar model of 
EM, the complex and frequency-dependent 
dielectric function ${\tilde \rho}(x,\omega)$
must satisfy the usual Kramers-Kronig dispersion 
relation \cite{lly2}.

To apply to cavity QED,
it is necessary to (a) go from free fields to 
interacting fields, and (b) go from classical fields to quantized fields.  Here we indicate
the rudiments of this program.

Consider a nonlinear wave equation: 
$D \Phi +  \lambda(x) \Phi^3 = 0$
where $\lambda(x)$ has support only within ${\cal R}$
(interaction occurs only inside the cavity).
Expanding this in terms of the eigenfunctions of noninteracting theory,
$( d/dt + i\omega_j) a_j = 
-i/(2\omega_j) \sum_{klm} \lambda_{jklm} a_k a_l a_m$
where
$\lambda_{jklm} = \int_0^{a^+}  \lambda f_j f_k f_l f_m dx$ 
Time-dependent perturbation of the linear theory (i.e., adding a time-dependent 
piece to $\rho(x)$) can be handled in very much the same way.
The point is that the dynamics is now expressed through {\em discrete}
variables $a_j(t)$

Since $\phi(x)$ and ${\hat \phi}(x)$ satisfy  
$\left[ \phi(x) , {\hat \phi}(y) \right] = i\delta (x-y)$ at equal times,
one obtains from (\ref{eq:aj}) commutation 
relations for $[a_j, a_k^+]$.  The commutators differ slightly
from $\delta_{ij}$, the difference being a measure of the dissipation.
Feynman rules can be developed in 
the standard way.
This description,
unlike the MOU approach, will be particularly
useful when the cavity is only slightly
leaky and the connection with a totally enclosed cavity is to
be emphasized.  Applications to emission, absorption and other
quantum phenomena are being developed. 
An important result will be to put idea of \citeasnoun{purcell} on resonance enhancement
of decay rates on a firm footing, by writing it as the contribution of one discrete QNM,
in a precise manner.
 % Time-dependent problems --- label{sect:timedep}
%========================================================== 
%file sect5
 
%========================================================== 
\section{Non-resonant contributions} 
\label{sect:gen} 
 
We have so far concentrated on the baseline
case in which the discontinuity and ``no tail"
conditions are satisfied, and the QNMs are complete.
However, in other cases there are non-resonant
contributions, and a brief discussion is given in this
Section.

Gravitational radiation from Schwarzschild black holes is described
by the KG equation; the potential $V(x)$,
related to the background metric, is continuous and  
goes asymptotically as $x^{-3} \log x$ 
as $x \rightarrow +\infty$ (apart from a centrifugal barrier),
violating both the discontinuity and the ``no tail" condition.   We indicate  
briefly how this situation affects the dynamics \cite{taillett,tail}. 
 
First, because there is no discontinuity,
there is a prompt signal due to the contribution
from the large semicircle. 
Secondly, the auxiliary function $g(\omega,x)$ 
has to be integrated from $x \rightarrow +\infty$. 
This leads to 
a cut on the negative $\mbox{Im $\omega$}$ axis, controlled 
by the spatial asymptotics of $V(x)$. 
The cut extends all the way to $\omega=0$, and determines the late time 
dynamics.  Generically, if $V(x) \sim x^{-\alpha} (\log x)^{\beta}$,
then $\phi(x,t) \sim t^{-\mu} (\log t)^{\nu}$, where $\mu$ and $\nu$ are determined by 
$\alpha$, $\beta$ and the angular momentum $l$. 
For example, for $\beta = 0$ or $1$,  
in general $\mu = 2l+\alpha$ and $\nu = \beta$.
It turns out, interestingly, 
that the case of the Schwarzschild black hole ($\alpha = 3$, $\beta=1$) 
is exceptional for $l > 0$: the leading term 
$t^{-(2l+\alpha)}\log t$ vanishes, and the dynamics is dominated by the 
next leading term $t^{-(2l+\alpha)}$.
Figure 2 shows precisely such a power-law tail.
However, the signal at intermediate times would still be dominated 
by the sum over QNMs, so that in practice, a restricted and approximate 
notion of completeness still holds. 

For the Schr\"{o}dinger equation,
outgoing waves are defined 
by $\phi(x) \sim e^{ikx}$ as $x \rightarrow \infty$, where
$k = \sqrt{2\omega}$ instead of $k = \omega$,
and moreover $k>0$.
Thus a continuum contribution cannot be avoided, and
the system decays 
by a power-law $t^{-\alpha}$ at large times,
due to the contribution near threshold ($k \approx 0$).
This phenomena was first noted by 
\citeasnoun{khafin} and \citeasnoun{winter}.
In particular, \citeasnoun{winter} considered a
1-d well bounded by an impenetrable wall and a delta-function 
potential and showed that the wavefunction inside the well decays as 
$t^{-3/2}$ as large $t$.  However, if the potential
is unbounded from below, no finite threshold
exists, and the large $t$ behavior is again given by the QNMs
\cite{sy1}.
This turns out to apply to some models of mini-superspace
(in which $x$ is essentially the scale factor of the universe) \cite{sy2}, and 
there is the intriguing possibility
that wavefunction of the universe at the end of the quantum era
is given by the lowest QNM, independent of
cosmological initial conditions.

The power-law decay in the case of the
Schr\"odinger equation comes from the threshold, whereas the
power-law in the KG case comes from the spatial tail of $V(x)$;
the origins of the two are entirely different.

Despite these complications, in time-independent problems, 
one is free to relabel $\omega^2 \mapsto \omega$, so the Schr\"{o}dinger case should not 
be different.  An application in this restricted time-independent 
sense is given in \citeasnoun{per4}. 

 % Generalizations --- \label{sect:gen}

%========================================================== 
%file sect 6
 
%========================================================== 
\section{Conclusion} 
\label{sect:diss} 

We have considered {\em linear}, {\em classical \/} waves 
in an {\em open} system of a certain type, 
and sketched a formalism in terms of an
QNM expansion, in a manner analogous to 
the description of conservative systems by its NMs. 
Dissipation is contained in these discrete QNMs themselves, 
especially 
in Im $\omega_j$.  Remarkably, apart from this feature, 
almost nothing needs to be changed, and the familiar tools
for hermitian systems can be carried over. 
These results are nontrivial, in that if either the discontinuity condition 
or the ``no tail" condition is violated, the QNMs would not be complete. 
The most significant feature in these cases is a power-law tail in
the large $t$ behavior.

With these mathematical concepts in place, the way is now open 
for formulating QED phenomena in an open cavity, using a {\em discrete} 
basis.
 % Discussion --- \label{sect:diss}

%==========================================================
\acknowledgments

This work is supported in part by a grant from the Hong Kong Research
Grants Council (Grant No. 452/95).  
The work of W. M. S. is supported by
U. S. N. S. F. Grant No. PHY 96-00507, NASA NCCS 5-153,
and the Institute of Mathematical Sciences
of The Chinese University of Hong Kong.
The work reported here has benefited from discussions with 
C. K. Au, R. K. Chang, H. Dekker, H. M. Lai, S. Y. Liu, A. Maassen van den Brink, 
R. Price, C. P. Sun and L. Yu.
We thank A. J. Campillo and J. D. Eversole for providing Figure 1b, and for
permission to use it.

%==========================================================

%==========================================================
%file ref

 %References

%==========================================================
\centerline{{\bf FIGURE  CAPTIONS} }

\begin{description}

\item[Fig. 1] (a) Schematic diagram of the EM spectrum observed outside
a linear optical cavity of length $a$, due to emission by a broadband
source inside the cavity.
(b) Experimental fluorescent spectra (upper) of a dye-doped ethanol
microsphere. Resonance peaks are found to correlate well with the
computed resonant mode positions (lower). Mode positions are plotted as
half arrows in a horizontal plane and projected by broken lines.
[Figure reproduced with permission from Fig. 5 of Chapter 4 of 
{\it Optical Processes in Microcavities} ed. R. K. Chang and  A. J.
Campillo, World Scientific (1996).]
 
\item[Fig. 2] Numerical simulation of the a
mplitude of linearized gravitational waves propagating on
a static, spherically symmetric blackhole background as a function of
time. Note particularly the domination by a ringing signal due to the
QNMs at intermediate times.

\item[Fig. 3] A spacetime diagram illustrating heuristically the
three different contributions to the Green's function.
Rays $1a$ and $1b$ show ``direct" propagation without scattering, 
and correspond to the prompt response. 
Ray $2$ suffers repeated scatterings at finite $x'$. 
For ``correct" frequencies, such multiple reflections add  
coherently so that retention of the wave is maximum. The amplitude 
of the wave decreases in a geometric progression with the 
number of scatterings, and thus decreases exponentially in time. 
These are QNM contributions. 
For a wave from a source point $y$ reaching a distant 
observer at $x$, there could be reflections at very large $x'$ by
$V(x')$, as indicated by ray $3$. For a general class of $V$, such ray
contributes to the late-time tail.

\item[Fig. 4] Positions of the QNMs for the dielectric rod model with
a small perturbation (crosses) compared to the unperturbed ones
(triangles).  First-order (squares) and second-order (circles)
perturbative results are also shown.

\item[Fig. 5] The locus of the lowest QNM for $l=1$ scalar waves
propagating on a Schwarzschild background, as the shell of magnitude
$\mu = 0.01$ is placed at different positions measured by $r_s$.
The solid line shows the
exact results, and the dashed line shows the first-order perturbative
results.  The square corresponds to a shell position of
$r_s/M_a = 2.22$ (the extreme value that satisfies the dominant
energy condition).  The triangles (on the solid line) and the crosses
(on the dashed line) refer to $r_s/M_a$ from 6 to 60 at intervals of 6.
The first-order result is seen to capture the qualitative features.

\end{description}

%==========================================================
\end{multicols}

\end{document}